\pdfoutput=1
\documentclass[superscriptaddress,nofootinbib,prl,two column]{revtex4}

\usepackage[colorlinks=true, pdfstartview=FitV, linkcolor=red, citecolor=blue,
urlcolor=blue]{hyperref}
\usepackage[dvipdfmx]{graphicx}
\usepackage{amsmath,amssymb}
\usepackage{color}
\usepackage{slashed}
\usepackage[normalem]{ulem}

\newcommand\m{\mu}
\newcommand\n{\nu}

\renewcommand\O{\Omega}

\newcommand{\Lag}{\mathcal{L}}

\newcommand{\ud}{\mathrm{d}}

\newcommand{\ue}{\mathrm{e}}

\renewcommand{\part}{{\rm part}}

\graphicspath{{./figures/}}

\begin{document}

\title{Fluctuations and correlations of quark spin in hot and dense QCD matter}

\author{Hao-Lei Chen}
\email{hlchen15@fudan.edu.cn}
\affiliation{Key Laboratory of Nuclear Physics and Ion-beam Application (MOE), Fudan University, Shanghai 200433, China}
\affiliation{Shanghai Research Center for Theoretical Nuclear Physics, NSFC and Fudan University, Shanghai 200438, China}
\author{Wei-jie Fu}
\email{wjfu@dlut.edu.cn}
\affiliation{School of Physics, Dalian University of Technology, Dalian 116024, China}
\affiliation{Shanghai Research Center for Theoretical Nuclear Physics, NSFC and Fudan University, Shanghai 200438, China}
\author{Xu-Guang Huang}
\email{huangxuguang@fudan.edu.cn}
\affiliation{Physics Department and Center for Particle Physics and Field Theory, Fudan University, Shanghai 200438, China}
\affiliation{Key Laboratory of Nuclear Physics and Ion-beam Application (MOE), Fudan University, Shanghai 200433, China}
\affiliation{Shanghai Research Center for Theoretical Nuclear Physics, NSFC and Fudan University, Shanghai 200438, China}

\author{Guo-Liang Ma}
\email{glma@fudan.edu.cn}
\affiliation{Key Laboratory of Nuclear Physics and Ion-beam Application (MOE), Fudan University, Shanghai 200433, China}
\affiliation{Shanghai Research Center for Theoretical Nuclear Physics, NSFC and Fudan University, Shanghai 200438, China}

\begin{abstract}

In this work, we examine the impact of QCD phase transitions on the quark spin fluctuations and correlations. We propose the quark-antiquark correlation, which relates to the vector meson spin alignment and the $\Lambda-\bar\Lambda$ correlation, can be used as a novel probe of the critical end point (CEP) in the QCD phase diagram. Using the Nambu-Jona-Lanisio model, we qualitatively study the properties of quark-antiquark spin correlations. Our findings reveal a peak structure near the CEP of the chiral phase transition, which may serve as an experimental signature of the CEP and account for the non-monotonic behavior of $\phi$ meson alignment at low collision energies observed recently in experiments.

\end{abstract}

\begin{titlepage}
\maketitle
\end{titlepage}


\paragraph*{Introduction.}

The large orbital angular momentum in the initial state of heavy ion collision experiments can result in rapid local rotation and the creation of fluid vortices~\cite{Betz:2007kg,Jiang:2016woz,Deng:2016gyh,Deng:2020ygd}. Such rotating matter has recently attracted significant attention. One particularly intriguing phenomenon is spin polarization and alignment of hadrons in heavy ion collisions, which has been observed experimentally~\cite{STAR:2017ckg,STAR:2022fan}. Theoretical predictions of global hyperon spin polarization~\cite{Liang:2004ph} have been successfully described by numerous subsequent theoretical frameworks~\cite{Becattini:2013fla,Fang:2016vpj,Karpenko:2016jyx,Becattini:2016gvu,Xie:2017upb,Li:2017slc,Shi:2017wpk,Sun:2018bjl,Xia:2018tes,Wei:2018zfb,Vitiuk:2019rfv,Li:2021zwq,Deng:2021miw,Guo:2021udq,Alzhrani:2022dpi,Wu:2022mkr}. For detailed reviews, see~\cite{Liang:2019clf,Gao:2020lxh,Huang:2020dtn,Liu:2020ymh,Becattini:2022zvf,Becattini:2024uha,Niida:2024ntm,Chen:2024afy}.

However, the experimental measurements of $\phi$ meson spin alignment~\cite{STAR:2022fan} significantly exceed theoretical expectations~\cite{Liang:2004xn}. Various mechanisms have been proposed to explain these measurements, including quark coalescence and fragmentation effects, the vorticity tensor, strong mesonic fields,  local spin alignment, medium-modified meson spectra,  turbulent color fields and viscous corrections~\cite{Yang:2017sdk,Sheng:2020ghv,Xia:2020tyd,Gao:2021rom,Muller:2021hpe,Li:2022vmb,Wagner:2022gza,Kumar:2023ghs,Li:2023tsf,Fang:2023bbw,Yin:2024dnu,Sheng:2024kgg,Lv:2024uev,Grossi:2024pyh,Chen:2024zwk}. However, most of these mechanisms predict the $\phi$ meson $\rho_{00}$ (the $00$-component of spin density matrix) to be less than $1/3$ which contradicts the experimental measurement. A proposed mechanism for strange quark polarization, based on the effective $\phi$ meson field and its local spatial and temporal correlation effects, suggests a higher $\phi$ meson $\rho_{00}$~\cite{Sheng:2019kmk,Sheng:2022wsy,Chen:2023hnb}. Additionally, experimental data also show non-monotonic behavior of $\phi$-meson $\rho_{00}$ at certain collision energies, suggesting the presence of unknown mechanisms.

Searching for the critical end point (CEP) is a key task in studying the quantum chromodynamics (QCD) phase structure~\cite{Luo:2017faz, Arslandok:2023utm}, and estimates of the CEP in the QCD phase diagram have arrived at convergent results from different approaches in recent years \cite{Fu:2019hdw, Gao:2020fbl, Gunkel:2021oya, Clarke:2024ugt}. In thermal equilibrium, a QCD system can be characterized by its dimensionless pressure, which is the logarithm of the QCD partition function. The Taylor expansion coefficients of this logarithm represent susceptibilities of conserved charges, defined as the $n$-th order derivative of pressure with respect to the chemical potential of the conserved charge. These susceptibility ratios can be experimentally measured by the cumulant ratios of the event-by-event conserved charge multiplicity distributions~\cite{Ding:2015ona}. Near the CEP, large correlation lengths make fluctuations of conserved charges sensitive probes for CEP detection~\cite{Stephanov:2011pb}. Recent results from the STAR collaboration on the ratio of fourth-order to second-order net-proton cumulants show a non-monotonic energy dependence with a significance of 3.1$\sigma$, suggesting the created system may have passed near the CEP~\cite{STAR:2021iop}.

In this work, we propose that quark spin correlation and fluctuation will be enhanced near the CEP, leading to observable imprints on vector meson spin alignment and $\Lambda-\bar\Lambda$ spin correlation \cite{Pang:2016igs,Du:2008zzb} in experiments. Our results reveal a deep connection between spin polarization and QCD phase transitions, which are often considered separately in the literature. Recent studies indicate that phase transitions can slightly affect average spin polarization~\cite{Sun:2024anu,Xu:2022hql,Singh:2021yba}, but they do not consider quark spin fluctuation. Inspired by baryon number fluctuation near the CEP, we anticipate that critical quark spin fluctuation can provide measurable effects. Using the NJL model, we qualitatively study quark-antiquark correlation and spin fluctuation near the CEP, finding a peak structure close to the CEP. We propose that along the freeze-out line, the non-monotonic behavior of hyperon correlation can serve as a signature for the CEP, alongside net baryon number fluctuation.


\paragraph*{Quark-antiquark spin correlation.}

We define the quark-antiquark spin correlation as
\begin{equation}\label{eq:PqPqbarC}
	\langle  P_{q}P_{\bar q}\rangle_c=\frac{4(\langle  S_{q}S_{\bar q}\rangle-\langle S_q\rangle\langle S_{\bar q}\rangle)}{\langle  N_{q}N_{\bar q}\rangle-\langle N_q\rangle\langle N_{\bar q}\rangle},
\end{equation}
where $S_{q}$ and $S_{\bar q}$ are the spins of the quark and antiquark, $N_{q}$ and $N_{\bar q}$ are their respective particle numbers. Since the quark has spin $1/2$,  a factor 4 in the numerator is introduced so that Eq.~(\ref{eq:PqPqbarC}) corresponds to the correlation of spin polarization rate. To study how this observable is affected by the CEP, we adopt the  Nambu-Jona-Lasinio (NJL) model under rotation with the  Lagrangian
\begin{equation}\label{eq:LagNJL}
	\Lag=\bar\psi[ie^\mu_a\gamma^a\nabla_\mu-m_0+\mu\gamma^0]\psi+G[(\bar\psi\psi)^2+(\bar\psi i\gamma^5{\vec\tau}\psi)^2],
\end{equation}
where $\psi = (u,d)^T$ is the two flavor quark field, $\nabla_\m = \partial_\mu +\Gamma_\mu$ the covariant derivative,  $\mu$ the quark chemical potential, related to the baryon chemical potential $\mu_B$ via $\mu=\mu_B/3$. The spin connection $\Gamma_\mu$ and vierbein $e^\mu_a$ are the same as \cite{Jiang:2016wvv,Chen:2015hfc}. In this work, we consider a rotating frame with a constant angular velocity $\vec \Omega=\Omega\hat z$, which can be described by the metric tensor
  \begin{equation}
  g_{\m\n}=
	\begin{pmatrix}
		1-r^2\O^2 &\O y& -\O x& 0 \\
       \O y & -1 & 0 & 0 \\
       -\O x & 0 & -1 & 0 \\
       0 & 0 & 0 & -1 \\	
	\end{pmatrix}
\end{equation}
with $r^2 = x^2 + y^2$. In the Lagrangian (\ref{eq:LagNJL}), we only take into account scalar interaction for simplicity since we are interested in the physics near the CEP qualitatively which is not sensitive to the microscopic interaction.
In general, the conserved quantity of the Lagrangian is total angular momentum along the rotation axis $J_z=L_z+S_z$. However, at $r=0$, only the spin contribution survives, allowing us to approximate the mean net quark spin by taking the derivative of the partition function
\begin{equation}
	\langle J_z\rangle{\Big |}_{r\approx 0}=\frac{\partial\ln Z}{\partial(\frac{\Omega}{T})}{\Big |}_{r\approx 0}\approx \frac{\partial\ln Z_0}{\partial(\frac{\Omega}{T})}=\langle S_z\rangle,
\end{equation}
where
\begin{equation}\label{eq:partition}
	\ln Z_0=-\frac{V}{T}V^0_{eff},
\end{equation}
and $V^0_{eff}$ is the local thermodynamic potential at $r=0$. We can further define the $n$-th order cumulant of spin fluctuation as
\begin{equation}\label{eq:cumulants}
 	C_n=\frac{\partial^n\ln Z_0}{\partial(\frac{\Omega}{T})^n}=-\frac{V}{T}\frac{\partial^n V^0_{eff}}{\partial(\frac{\Omega}{T})^n}.
 \end{equation}
 The second order cumulants corresponds to the variance $\langle \delta S^2\rangle$ which is relevant in this work.

The thermodynamic potential can be directly calculated \cite{Jiang:2016wvv,Chen:2015hfc}. In order to obtain the correlation of quark-antiquark, we modified the thermodynamic potential to be
\begin{widetext}

\begin{equation}\label{eq:Veff0}
\begin{split}
V_{\mathrm{eff}}^0(\Omega_q,\Omega_{\bar q}, \mu_q,\mu_{\bar q},\mu)=&\frac{(m-m_0)^2}{4G}-N_cN_f\int_0^\Lambda\frac{\ud^3p}{(2\pi)^3}[2\varepsilon_p
-F(\Omega_q)-F(-\Omega_q)
-\bar F(\Omega_{\bar q})-\bar F(-\Omega_{\bar q})],
\end{split}
\end{equation}
\end{widetext}
with
\begin{equation} 
\begin{split}
F(\Omega_q)&=T\ln(1+\ue^{-(\varepsilon_p-\mu-\mu_q-\Omega_q/2)/T}),\\
\bar F(\Omega_{\bar q})&=T\ln(1+\ue^{-(\varepsilon_p+\mu-\mu_{\bar q}-\Omega_{\bar q}/2)/T}),
\end{split}
\end{equation}
and the dispersion relation $\varepsilon_p=\sqrt{p^2+m^2}$. Here, $\Omega_q$ and $\Omega_{\bar q}$ are rotation acting only on quark and antiquark, respectively, and we also introduce a particle number chemical potential $\mu_q$ for quark and $\mu_{\bar q}$ for antiquark. A detailed derivation of Eq.~(\ref{eq:Veff0}) is presented in the supplemental materials.
Then by taking derivative we can obtain the variance of quark and antiquark spin fluctuation
\begin{equation}
	\begin{split}
		\langle \delta S_{q}^2\rangle=&-\frac{V}{T}\frac{\partial^2V_{eff}(\Omega_q,\Omega_{\bar q})}{\partial(\frac{\Omega_q}{T})^2}{\Big |}_{\Omega_q=\Omega_{\bar q}=\Omega}, \\
		\langle \delta S_{\bar q}^2\rangle=&-\frac{V}{T}\frac{\partial^2V_{eff}(\Omega_q,\Omega_{\bar q})}{\partial(\frac{\Omega_{\bar q}}{T})^2}{\Big |}_{\Omega_q=\Omega_{\bar q}=\Omega},
	\end{split}
\end{equation}
and the variance of quark and antiquark particle number
\begin{equation}
	\begin{split}
		\langle \delta N_{q}^2\rangle=&-\frac{V}{T}\frac{\partial^2V_{eff}}{\partial(\frac{\mu_{ q}}{T})^2}{\Big |}_{\mu_q=\mu_{\bar q}=0}, \\
		\langle \delta N_{\bar q}^2\rangle=&-\frac{V}{T}\frac{\partial^2V_{eff}}{\partial(\frac{\mu_{\bar q}}{T})^2}{\Big |}_{\mu_q=\mu_{\bar q}=0}.
	\end{split}
\end{equation}

By separating total spin into quark and antiquark part as $S=S_q+S_{\bar q}$, it is easy to prove the relation
\begin{equation}
\begin{split}
	\langle  S_{q}S_{\bar q}\rangle-\langle S_q\rangle\langle S_{\bar q}\rangle=\frac{1}{2}(\langle \delta S^2\rangle-\langle \delta S_{q}^2\rangle-\langle \delta S_{\bar q}^2\rangle),
 \end{split}
\end{equation}
and similarly
\begin{equation}
	\langle  N_{q}N_{\bar q}\rangle-\langle N_q\rangle\langle N_{\bar q}\rangle=\frac{1}{2}(\langle \delta N^2\rangle-\langle \delta N_{q}^2\rangle-\langle \delta N_{\bar q}^2\rangle).
\end{equation}
Finally, we obtain
\begin{equation}
	\langle  P_{q}P_{\bar q}\rangle_c=\frac{4(\langle \delta S^2\rangle-\langle \delta S_{q}^2\rangle-\langle \delta S_{\bar q}^2\rangle)}{\langle \delta N^2\rangle-\langle \delta N_{q}^2\rangle-\langle \delta N_{\bar q}^2\rangle}.
\end{equation}
Note that although the factor $V/T$ appears in Eq.~(\ref{eq:partition}), it will be canceled in the ratio, thus $\langle  P_{q}P_{\bar q}\rangle_c$ is not affected by the system volume.


\paragraph*{Numerical results.}\label{sec:corr}

For numerical calculation, we choose the parameters listed in Table.~\ref{table:para1} for 2 flavor NJL model \cite{Buballa:2003qv}.
\begin{table}[htbp]
\centering
\caption{Parameter set used for the NJL model}
\label{table:para1}
\begin{ruledtabular}
\begin{tabular}{ccccc}
$\Lambda$ [MeV]&  $m_0$ [MeV] & {}$G\Lambda^2$ & $N_c$& $N_f$\\ \hline
$587.9$  & $5.6$ & $2.44$ & $3$& $2$\\ 
\end{tabular}
\end{ruledtabular}
\end{table}
In this case the CEP is located at $T_{CEP}=82.1$ MeV and $\mu_{CEP}=322$ MeV. Here we fix the angular velocity to be $\Omega=10$MeV, which is a typical value for the fluid vorticity in the experiment.
For comparison, we also calculate the correlation of quark-antiquark spin without including critical fluctuation, which can be easily written down with distribution function $f_q$ and $f_{\bar q}$ as
\begin{equation}\label{eq:estimate}
	\langle  P_{q}P_{\bar q}\rangle_{0}=\frac{\int\ud^3 p (f_q^\uparrow-f_q^\downarrow)(f_{\bar q}^\uparrow-f_{\bar q}^\downarrow)}{\int\ud^3 p (f_q^\uparrow+f_q^\downarrow)(f_{\bar q}^\uparrow+f_{\bar q}^\downarrow)}.
\end{equation}
where 
\begin{equation}
	f_q^{\uparrow/\downarrow}=\frac{1}{\ue^{\beta(\epsilon_p-\mu\mp\frac{\Omega}{2})}+1},\quad f_{\bar q}^{\uparrow/\downarrow}=\frac{1}{\ue^{\beta(\epsilon_p+\mu\mp\frac{\Omega}{2})}+1},
\end{equation}
are the distributions of spin-up and spin-down quarks and anti-quarks. We assume the distribution functions are homogeneous in space, which simplifies our analysis while still capturing the essential physics of the system.

%
\begin{figure}[t]
\includegraphics[width=1\columnwidth]{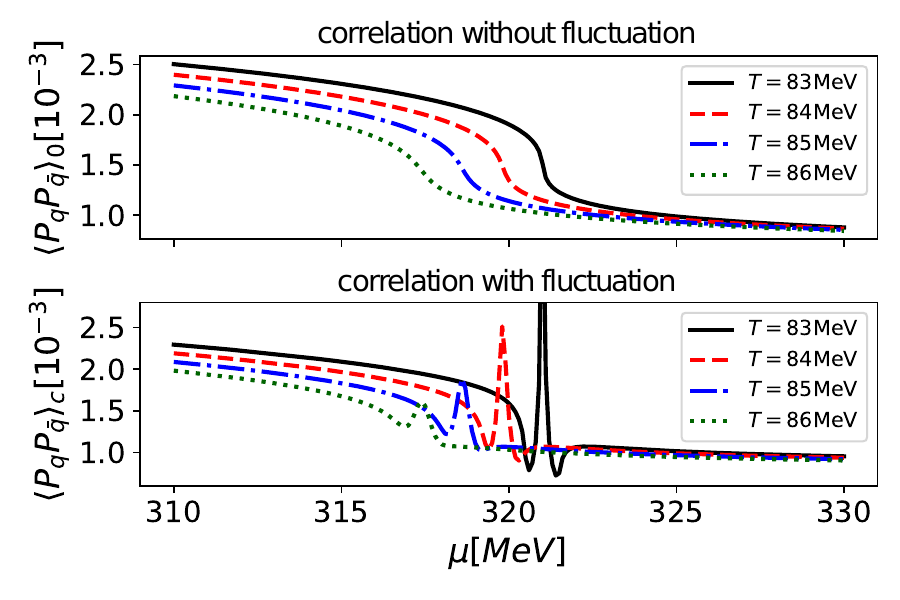}
    \caption{Upper panel: Quark-antiquark spin correlation $\langle  P_{q}P_{\bar q}\rangle_{0}$ without critical fluctuation as a function of chemical potential. Lower panel: Quark-antiquark spin correlation $\langle  P_{q}P_{\bar q}\rangle_{c}$ with critical fluctuation as a function of chemical potential at $r=0$. The angular velocity is chosen to be $\Omega=10$MeV.}
    \label{fig:PqPqbarC0}
\end{figure}
%

\begin{figure}[t]
\includegraphics[width=1\columnwidth]{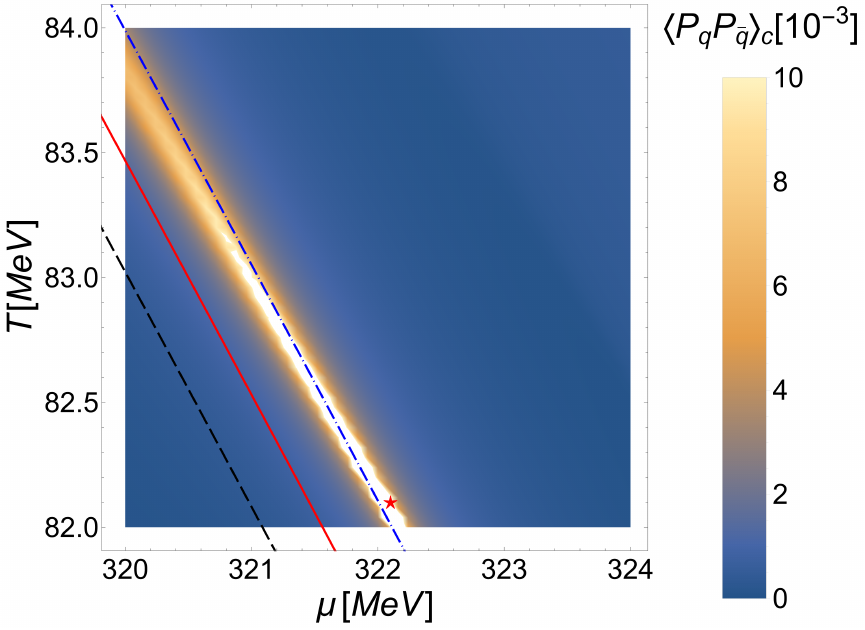}
    \caption{The density plot of the connected correlation $\langle  P_{q}P_{\bar q}\rangle_c$  at $r=0$ with $\Omega=10$ MeV. The solid, dash and dot-dash line represent the hypothetical freezout line-1, 2, and 3 in Table~\ref{table:linepara}, respectively. The red star stands for the CEP.}
    \label{fig:DensityPlot}
\end{figure}

The numerical results of $\langle  P_{q}P_{\bar q}\rangle_{0}$ and $\langle  P_{q}P_{\bar q}\rangle_{c}$ near the CEP are shown in Fig.~\ref{fig:PqPqbarC0}.     Both the upper and lower panels show similar trends. However, $\langle  P_{q}P_{\bar q}\rangle_{c}$  has a peak structure which becomes higher as we decrease the temperature and approach the CEP. Besides, we also observe dip structures on the two sides of the peak when approaching CEP (which is most obvious at $T=83$MeV in Fig.~\ref{fig:PqPqbarC0}). These peak  and dip structures  indicate the appearance of critical fluctuations near the CEP. Fig.~\ref{fig:DensityPlot} presents the density plot of the connected correlation $\langle  P_{q}P_{\bar q}\rangle_c$ and three freezeout lines. Note that the phase boundary obtained in the current NJL model is not in quantitative agreement with QCD results \cite{Bellwied:2015rza, HotQCD:2018pds, Fu:2019hdw, Fu:2022gou}. Instead of using more realistic freezeout lines, e.g., those extracted from experimental measurements \cite{STAR:2017sal, Andronic:2017pug, Fu:2021oaw}, we employ hypothetical freezeout lines, not far away from the CEP in this NJL model, which are parameterized as the form
\begin{equation}\label{eq:freezeout}
	T = a_0 - a_2\mu^2 - a_4\mu^4\,.
\end{equation}
We utilize three distinct sets of parameters, as shown in Table.~\ref{table:linepara}.  From Fig.~\ref{fig:DensityPlot} one can see that the distances from the CEP to the three freezeout lines are different. When the freezeout line intersects the narrow crossover region near the CEP, the spin correlation will exhibit a noticeable peak structure.
%

%
\begin{table}[htbp]
\centering
\caption{Parameters for freezeout lines}
\label{table:linepara}
\tabcolsep=0.35cm
\begin{ruledtabular}
\begin{tabular}{cccc}
{} &  $a_0$[GeV] & $a_2$[GeV${}^{-1}$] & $a_4$[GeV${}^{-3}$]\\ \hline
freezeout-1  & 0.1831 & 0.4856 & 4.7589\\ 
freezeout-2  & 0.1831 & 0.49 & 4.7589\\
freezeout-3  & 0.1831 & 0.4756 & 4.8069\\ 
\end{tabular}
\end{ruledtabular}
\end{table}
%

\begin{figure}[t]
\includegraphics[width=1\columnwidth]{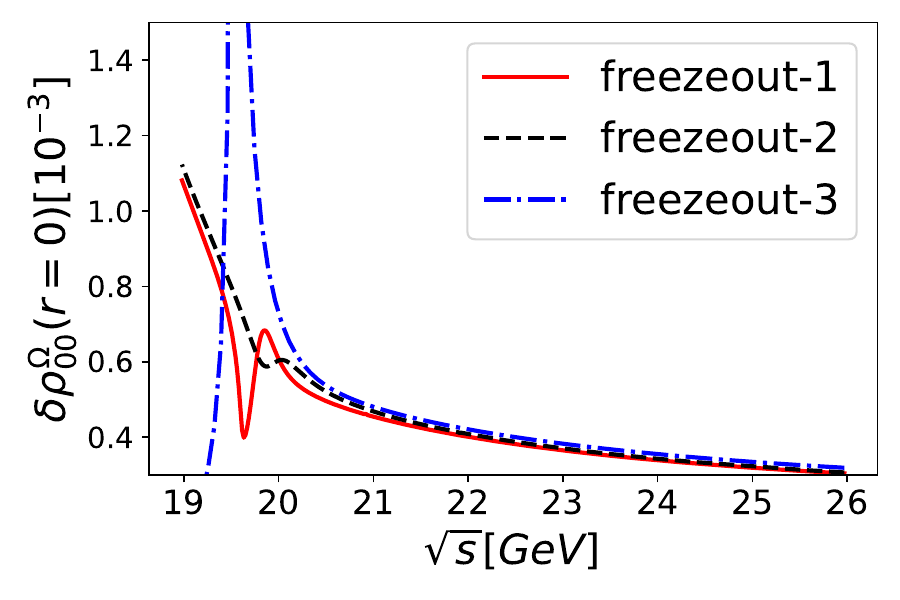}
    \caption{$\delta\rho_{00}^\Omega$ at $r=0$ as a function of the collision energy $\sqrt{s}$ along the three different freezeout lines in Table.~\ref{table:linepara}.}
    \label{fig:rho00}
\end{figure}

Next, to study qualitatively how these peak and dip structures manifest in the collision energy dependence of  vector meson spin alignment, we write the spin density matrix element $\rho_{00}$ as~\cite{Liang:2004xn}
\begin{equation}
    \rho_{00}=\bar\rho_{00}-\delta\rho_{00}^\Omega. \label{eq:rho00-compo}
\end{equation} 
where $\rho_{00}$ is decomposed into a sum of contribution from thermodynamic fluctuations $-\delta\rho_{00}^\Omega$ investigated in this work and that from other sources $\bar\rho_{00}$, e.g., the strong force field as suggested in \cite{Sheng:2022wsy}. The minus sign on the right side in Eq.~(\ref{eq:rho00-compo}) is introduced for convenience, since the thermodynamic spin correlation is negative. If we neglect any other contributions except the thermodynamic fluctuations, one arrives at
\begin{equation}
    \rho_{00}=\frac{1-\langle  P_{q}P_{\bar q}\rangle_c}{3+\langle  P_{q}P_{\bar q}\rangle_c}\approx \frac{1}{3}-\frac{4}{9}\langle  P_{q}P_{\bar q}\rangle_{c},
\end{equation} 
from which one obtains $\delta \rho_{00}^\Omega=\frac{4}{9}\langle  P_{q}P_{\bar q}\rangle_{c}$. Moreover, one has $\bar\rho_{00}=1/3$ if there is no other effects. Note that the decomposition in Eq.~(\ref{eq:rho00-compo}) as well as $\delta \rho_{00}^\Omega=\frac{4}{9}\langle  P_{q}P_{\bar q}\rangle_{c}$ is valid if the value of $\bar\rho_{00}$ is not far away from 1/3. In this work, we assume that the dominant contribution in $\bar\rho_{00}$  comes from the strong force field \cite{Sheng:2022wsy}, which leads to $\bar\rho_{00}>1/3$. In
Fig.~\ref{fig:rho00}, we present the contribution of critical fluctuation $\delta \rho_{00}^\Omega$ at $r=0$. Here we assume the quark chemical potential depends on the collision energy through $\mu=2.477/(1+0.343\sqrt{s})$~\cite{Begun:2016pdy}.
It is evident  that  $\delta\rho_{00}^\Omega>0$, which aligns with the expectation that rotation does not distinguish between quark and antiquark. Notice that the dot-dashed line (freezeout-3) is very close to the CEP, which exhibits a sharper peak and dip structure compared to the other lines.
This suggests that the contribution of critical fluctuations to the quark-antiquark spin correlation is significant and is potentially detectable in experiment. However, the peaks in our qualitative results are relatively narrow. In realistic situation, we expect these peaks to be wider and higher, as we will discuss further.

%
%

%
\begin{figure}[t]
\includegraphics[width=1\columnwidth]{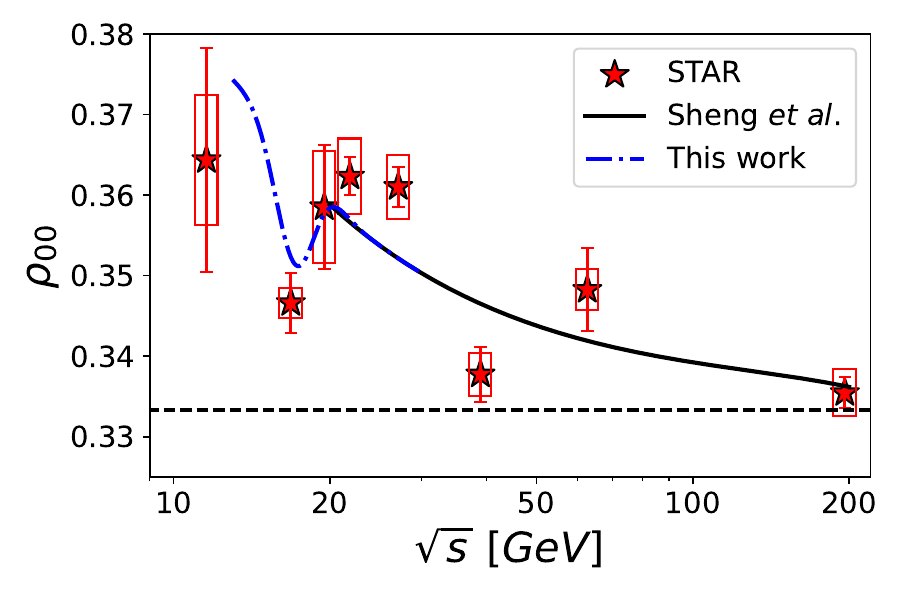}
    \caption{A schematic figure including the critical fluctuation. The dashed line stands for $\rho_{00}=1/3$. The dashed dot line is based on the freezeout-3 and our argument. The solid line is from Ref. \cite{Sheng:2019kmk}. The STAR data is from Ref.~\cite{Wilks}.}
    \label{fig:exp}
\end{figure}
%

In this work, we use the NJL model near the rotating center (i.e. $r=0$) to conduct the calculations, which allows us to obtain the cumulants of spin by directly taking derivatives. As we move further from the center, orbital angular momentum becomes significant. In the supplemental materials the calculations are also extended to the case of $r\ne 0$, where the contribution of orbital angular momentum is taken into account. It is found that the spin correlations increase with $r$ and thus the behavior discussed here is even more prominent at $r\ne 0$. As $r$ increases, the rotational effect becomes stronger due to the contribution from the large orbital angular momentum. Consequently, the location of the CEP will shift from that at $r=0$. Previous model studies \cite{Wang:2018sur,Chen:2023cjt} indicate that both $T_{CEP}$ and $\mu_{CEP}$ change with increasing $r$ or $\Omega$. Considering this, the position of the peak will vary for different values of $r$, bringing the CEP closer to the freezeout line as shown in the supplemental materials. Summing all the peaks at different $r$ leads to a higher and wider overall peak structure. As a result, we can estimate the order of overall contribution from critical fluctuations $\delta \rho_{00}^\Omega$ to be about $10^{-2}$. To illustrate this concept, we present a schematic figure of $\phi$ meson spin alignment $\rho_{00}$ (Fig.~\ref{fig:exp}). In this figure, we incorporate the contribution of critical fluctuations into the results of \cite{Sheng:2022wsy}, which results in a significant dip in the transition region. Notably, recent results from BES-II also show a similar dip structure~\cite{Wilks}, which might support our proposal.
In conclusion, we suggest that the effects of critical fluctuations on quark-antiquark correlations are measurable in experiments and may help understand the non-monotonic behavior observed in current experimental data.


\paragraph*{Conclusions and discussions.}

In this study, we investigate critical quark spin fluctuations near the CEP, a phenomenon hitherto unexplored in the literature. Our analysis reveals a pronounced peak in quark spin correlations near the CEP, primarily driven by the rapid variation of the order parameter within the crossover region. This observation serves as a distinctive signature of approaching the CEP.

We propose that the quark-antiquark correlation is potentially measurable through phenomena such as the $\phi$ meson spin alignment or hyperon-anti-hyperon correlation~\cite{Lv:2024uev}
\begin{equation}
\frac{N^{\uparrow\uparrow}_{H\bar H}+N^{\downarrow\downarrow}_{H\bar H}-N^{\uparrow\downarrow}_{H\bar H}-N^{\downarrow\uparrow}_{H\bar H}}{N^{\uparrow\uparrow}_{H\bar H}+N^{\downarrow\downarrow}_{H\bar H}+N^{\uparrow\downarrow}_{H\bar H}+N^{\downarrow\uparrow}_{H\bar H}},
\end{equation}
which could be significantly influenced in the proximity of the CEP, where $N^{s_1s_2}_{H\bar H}$ denotes the number of hyperon-anti-hyperon pair with spin $s_1$ and $s_2$, respectively. While further experimental data with improved statistical significance is necessary to confirm this second-order correlation, we suggest it as a promising candidate for future observations.

We should mention that since  model calculations~\cite{Chen:2021aiq} and lattice studies~\cite{Braguta:2021jgn,Yang:2023vsw} on the phase transition line in $T$-$\Omega$ plane are in contradiction, the qualitative behavior of phase structure given by the NJL model may be not consistent with the one in the real world. However, the observed quark spin fluctuation near the CEP remains a robust conclusion, independent of whether the critical temperature increases or decreases with rotation.

We also emphasize that although in this work we use a 2-flavor NJL model which does not include the $s$ quark. We have also checked the case of 2+1 flavor NJL model, computed the spin fluctuation of $s$-quark, and have found that the qualitative conclusion does not change. In principle, our conclusions should apply to other vector mesons, such as the $K^*$ and $\rho$ mesons. However, the $K^*$ and $\rho$ mesons are more likely to be affected by hadronic interactions, which could wash out the spin correlation, in comparison to the $\phi$ mesons \cite{Shen:2021pds}. Thus we expect the $\phi$ spin alignment serves as an appropriate candidate for the measurement.Moreover, we have also checked the case of PNJL model in the supplemental materials, and found that inclusion of the gluon background field has a minor effect on our results.

Note that the quark coalescence scenario is assumed in our theoretical calculations, which is widely used in the studies of spin alignment of vector mesons and hyperon polarization \cite{Liang:2004ph, Sheng:2022wsy}. The spin density matrix of vector mesons or hyperons carries on the spin of constituent quarks. In the regime of low $p_T$ and central rapidity, where the experimental measurements receive the most contributions, the mechanism of quark coalescence plays the dominant role, which is also consistent with thermodynamic description of the bulk property. This underlies the survival of the quark spin fluctuations from the hadronization, that is similar with the case of experimental measurements of the net-proton fluctuations.

It is also worth noting that the spin obtained by taking derivatives with respect to $\Omega_q$ or $\Omega_{\bar q}$ corresponds to the canonical spin tensor, which is a natural and convenient choice in our framework. However, the spin correlations studied here may, in general, depend on the choice of the so-called pseudo-gauge~\cite{Becattini:2018duy}. In principle, one can define a spin potential related to a chosen spin tensor under a given pseudo-gauge and compute the corresponding expectation values of spin correlations, following a procedure similar to that used in our analysis. While such an extension lies beyond the scope of the present study, it remains an interesting direction for future study. Nevertheless, we expect the qualitative behavior of spin correlations near the CEP to remain robust, provided that a consistent pseudo-gauge is employed throughout the analysis, especially near $r=0$, as considered in this work.

The spin related effects such as the spin polarization and alignment are still  active topics in the field of high energy nuclear physics. In this work, we mainly focus on the qualitative behavior of spin fluctuations within the NJL model calculations at thermal equilibrium. More detailed and realistic studies, such as the effects of the magnetic field and viscosity during the dynamic evolution from the initial stage \cite{Sahoo:2023xnu, Huang:2011ru, Huang:2020dtn}, are required, which we leave for future studies.

\section*{Acknowledgment}

This work is supported by the National Natural Science Foundation of China under Grant Nos. 12325507, 12247133, 12175030, 12147101, 12075061, 12225502, the National Key Research and Development Program of China under Grant No. 2022YFA1604900, the Guangdong Major Project of Basic and Applied Basic Research under Grant No. 2020B0301030008, and the Natural Science Foundation of Shanghai under Grant No. 23JC1400200.

\bibliography{ref}


\clearpage

\title{Supplemental materials for ``Fluctuations and correlations of quark spin in hot and dense QCD matter''}
\begin{titlepage}
\maketitle
\end{titlepage}
\appendix

\onecolumngrid


\section{Thermodynamic potential with rotations in the NJL model}

In order to obtain the local thermodynamic potential in the NJL model within the mean field approximation, we assume $\partial_r m(r)\ll m^2$, that is, the local density approximation. A straightforward calculation leads to the $r$-dependent potential \cite{Jiang:2016wvv, Chen:2015hfc}, 
\begin{equation}\label{eq:Veff}
\begin{split}
\tiny V_{\mathrm{eff}}(r)=&\frac{(m-m_0)^2}{4G}-N_cN_f\sum_{l=-\infty}^\infty\int_0^\Lambda\frac{p_t\ud p_t\ud p_z}{(2\pi)^2}J_l^2(p_tr)\Big[2\varepsilon_p+T\ln(1+\ue^{-\beta(\varepsilon_p-\mu-\Omega l-\Omega/2)})\\&+T\ln(1+\ue^{-\beta(\varepsilon_p+\mu+\Omega l+\Omega/2)})+T\ln(1+\ue^{-\beta(\varepsilon_p-\mu-\Omega l+\Omega/2)})+T\ln(1+\ue^{-\beta(\varepsilon_p+\mu+\Omega l-\Omega/2)})\Big]
.
\end{split}
\end{equation}
 with $\varepsilon_p=\sqrt{p^2+m^2}$ the dispersion relation, $l$ the quantum number of angular momentum  and a hard cutoff $\Lambda$ in the momentum integration, where $p_z$ and $p_t$ stand for the longitudinal and transverse momentum, respectively. Note that we have shifted the summation of the angular momentum quantum number $l$ in comparison to Refs. \cite{Jiang:2016wvv, Chen:2015hfc}. We  consider slow rotations $\Omega\ll T$ and the physic near the system center.
Under such assumption, the  boundary condition is less important, and thus we can approximately write
\begin{equation}
    V_{\mathrm{eff}}(r)=V^0_{\mathrm{eff}}+O(\beta^4\Omega^2r^2), \label{eq:Veff-expa}
\end{equation}
where $V^0_{\mathrm{eff}}$ is the local thermodynamic potential at $r=0$. It is obvious that
only the mode with $l=0$ contributes to $V_{\mathrm{eff}}^0$, which means that only the spin contributes at $r=0$.   The orbital angular momentum contribution is $r$-dependent, which is high-order one and will be discussed in the following. The explicit expression of the leading term in Eq.~(\ref{eq:Veff-expa}) reads
\begin{equation}
\begin{split}
V_{\mathrm{eff}}^0(\Omega_q,\Omega_{\bar q}, \mu_q,\mu_{\bar q},\mu)=&\frac{(m-m_0)^2}{4G}-N_cN_f\int_0^\Lambda\frac{\ud^3p}{(2\pi)^3}2\varepsilon_p\\
&-N_cN_f\int_0^\infty\frac{\ud^3p}{(2\pi)^3}\Big[T\ln(1+\ue^{-(\varepsilon_p-\mu-\Omega_q/2-\mu_q)/T})+T\ln(1+\ue^{-(\varepsilon_p-\mu+\Omega_q/2-\mu_q)/T})\\
&+T\ln(1+\ue^{-(\varepsilon_p+\mu-\Omega_{\bar q}/2-\mu_{\bar q})/T})+T\ln(1+\ue^{-(\varepsilon_p+\mu+\Omega_{\bar q}/2-\mu_{\bar q})/T})\Big].
\end{split}
\end{equation}
Here we have already introduced the rotation and particle number chemical potential for quark and antiquark respectively. Then we can directly obtain the quark and antiquark spin
\begin{equation}
	\begin{split}
		\langle S_{q}\rangle=&-\frac{V}{T}\frac{\partial V_{\mathrm{eff}}}{\partial(\frac{\Omega_q}{T})}{\Big |}_{\substack{\Omega_q=\Omega_{\bar q}=\Omega\\
  \mu_q=\mu_{\bar q}=0}}, \quad
  \langle S_{\bar q}\rangle=-\frac{V}{T}\frac{\partial V_{\mathrm{eff}}}{\partial(\frac{\Omega_{\bar q}}{T})}{\Big |}_{\substack{\Omega_q=\Omega_{\bar q}=\Omega\\
  \mu_q=\mu_{\bar q}=0}},
	\end{split}
\end{equation}
the quark and antiquark particle number
\begin{equation}
	\begin{split}
		\langle N_{q}\rangle=&-\frac{V}{T}\frac{\partial V_{\mathrm{eff}}}{\partial(\frac{\mu_q}{T})}{\Big |}_{\substack{\Omega_q=\Omega_{\bar q}=\Omega\\
  \mu_q=\mu_{\bar q}=0}}, \quad
  \langle N_{\bar q}\rangle=-\frac{V}{T}\frac{\partial V_{\mathrm{eff}}}{\partial(\frac{\mu_{\bar q}}{T})}{\Big |}_{\substack{\Omega_q=\Omega_{\bar q}=\Omega\\
  \mu_q=\mu_{\bar q}=0}}.
	\end{split}
\end{equation}

At the end of this section, we make some comments on the validation of our assumption as follows .
As explained in \cite{Ayala:2021osy}, one can assume that the fermion is totally dragged by the vortical motion, i.e. $\theta+\Omega t=0$, where $\theta$ is the azimuthal angle in the cylindrical coordinate.  As a consequence, there is no contribution from the orbital angular momentum. Thus our assumption is equivalent to that in \cite{Ayala:2021osy}.

%
\begin{figure}[t]
\includegraphics[width=0.5\textwidth]{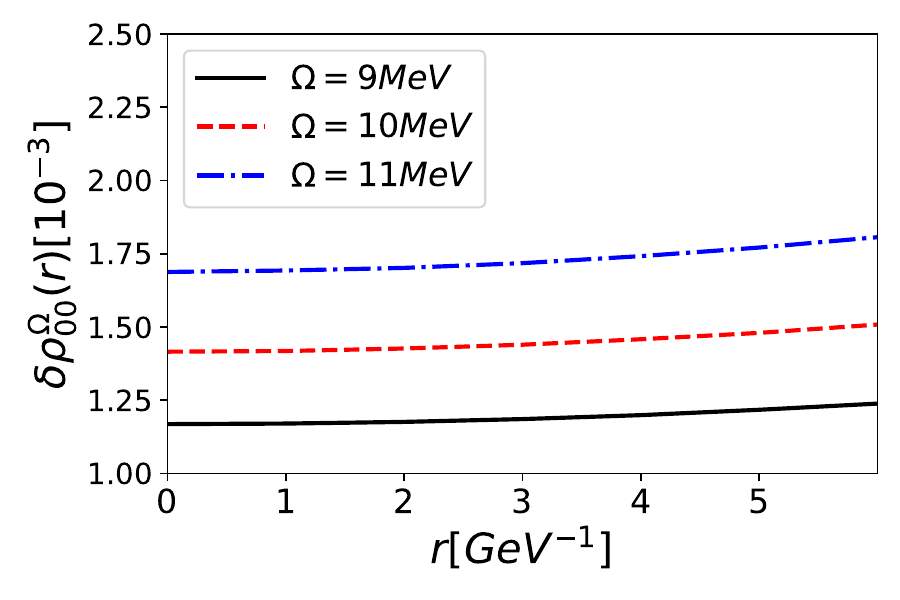}
\caption{Contribution of the thermodynamic fluctuations to the spin density matrix element $\delta\rho_{00}^\Omega$ as a function of the radial coordinate $r$ with several values of the angular velocity, where we have chosen a thermodynamic state in the phase diagram far away from the CEP with $T=85$ MeV and $\mu=250$ MeV.}
\label{fig:r-dep}
\end{figure}
%

%
\begin{figure}[t]
\includegraphics[width=0.5\textwidth]{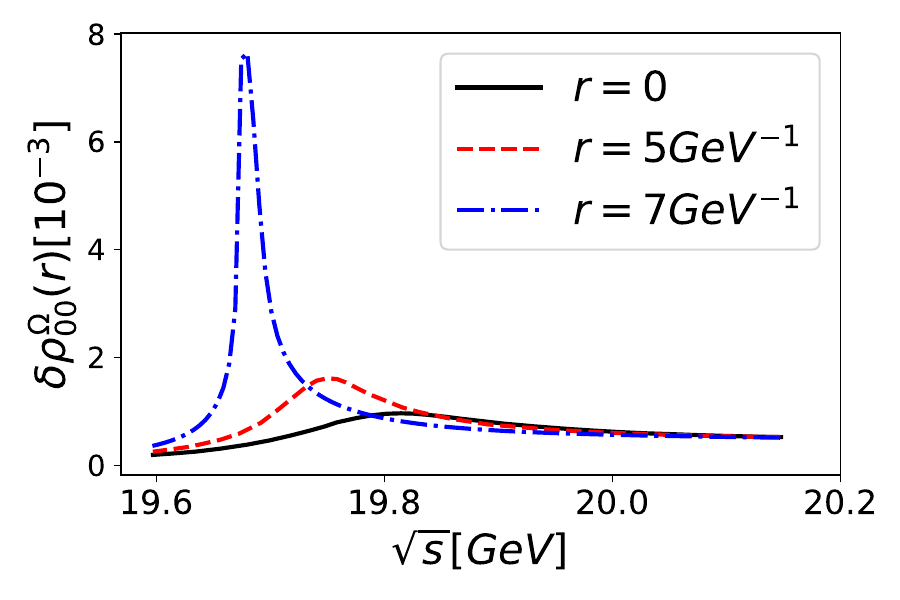}
\caption{Contribution of the thermodynamic fluctuations to the spin density matrix element $\delta\rho_{00}^\Omega$ as a function of the collision energy $\sqrt{s}$ along the freezeout line-2 in Table II as shown the main text, where the radial coordinate $r$ is varied and the angular velocity is chosen to be $\Omega=10$ MeV.}
\label{fig:oam}
\end{figure}
%


\section{Contribution of the orbital angular momentum at $r\neq 0$}

For $r\neq 0$, there is a contribution from the orbital angular momentum. To separate the spin and the orbital angular momentum, we rewrite the effective potential in Eq.~(\ref{eq:Veff}) as 
\begin{equation}\label{eq:Veff2}
\begin{split}
&V_{\mathrm{eff}}(\Omega_q^s,\Omega_{\bar q}^s,\Omega, \mu_q,\mu_{\bar q},\mu;r)\\
=&\frac{[m(r)-m_0]^2}{4G}-N_cN_f\int_0^\Lambda\frac{\ud^3p}{(2\pi)^3}2\varepsilon_p-\sum_{l=-\infty}^\infty N_cN_f\int_0^\infty\frac{\ud^3p}{(2\pi)^3}J_l^2(p_tr)\Big[T\ln(1+\ue^{-(\varepsilon_p-\mu-\Omega_q^s/2-\Omega l-\mu_q)/T})\\
&\hspace{-0.2cm}+T\ln(1+\ue^{-(\varepsilon_p-\mu+\Omega_q^s/2-\Omega l-\mu_q)/T})+T\ln(1+\ue^{-(\varepsilon_p+\mu-\Omega_{\bar q}^s/2+\Omega l-\mu_{\bar q})/T})+T\ln(1+\ue^{-(\varepsilon_p+\mu+\Omega_{\bar q}^s/2+\Omega l-\mu_{\bar q})/T})\Big],
\end{split}
\end{equation}
where $\Omega^s_q$ and $\Omega_{\bar q}^s$ stand for the angular velocities that only couple to the spin of quark and antiquark respectively.  Then we can numerically take derivative by $\Omega^s_q$ and $\Omega_{\bar q}^s$ to obtain the spin correlation similar as the $r=0$ case, while not entangled with the the angular momentum related term $\Omega\, l$. Certainly, $\Omega^s_q=\Omega_{\bar q}^s=\Omega$ is implemented after the differentiations.

The relevant numerical results are shown in Fig.~\ref{fig:r-dep}, where the contribution of the thermodynamic fluctuations to the spin density matrix element, i.e., the quark-antiquark spin correlation, are calculated with angular momenta taken into account in Eq.~(\ref{eq:Veff2}). In order to separate the direct effect, arising from the angular momentum, from the indirect one due to the change of phase structure with the variation of radial coordinate $r$, e.g., the movement of CEP, we have chosen a thermodynamic state in the phase diagram far away from the CEP with $T=85$ MeV and $\mu=250$ MeV on purpose. One can find that the spin correlation increases slowly with $r$. Moreover, we also perform the calculations on the freezeout line, and the obtained results are presented in Fig.~\ref{fig:oam}. One observes that the height of the peak is enhanced as we increase the distance from the center. This is mainly attributed to the fact that the CEP is drawn closer to the freezeout line by the stronger rotation effect at $r\ne 0$ in comparison to the case of $r=0$. Besides, the location of peak is also shifted due to the change of CEP location in the phase diagram. It should be mentioned that how exactly the peak is shifted (to higher or lower energy) might be model dependent, since typical model studies find the CEP moves along a certain trajectory as one changes the angular velocity \cite{Hernandez:2024nev, Singha:2024tpo, Morales-Tejera:2025qvh}.  

\begin{figure}[t]
\includegraphics[width=0.5\textwidth]{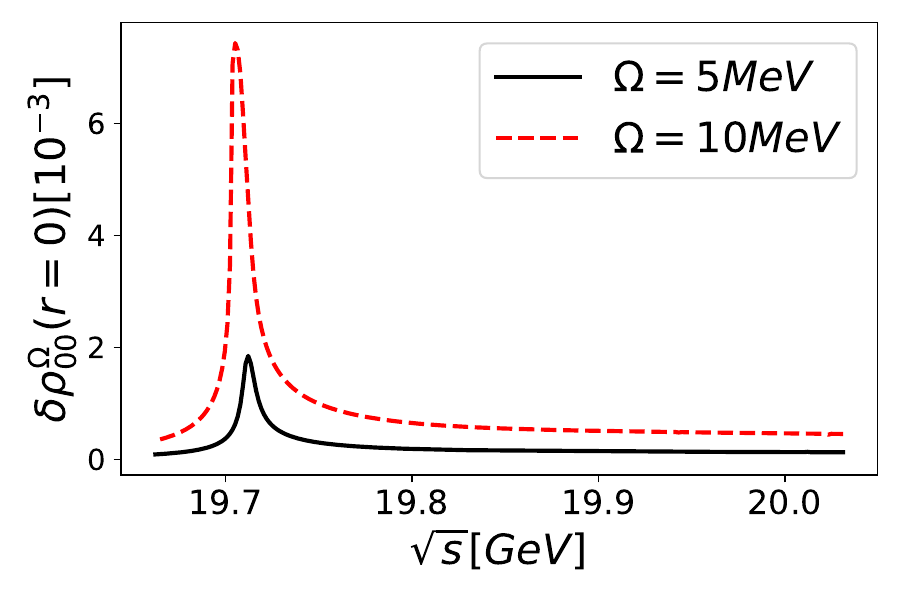}
\caption{Contribution of the thermodynamic fluctuations to the spin density matrix element $\delta\rho_{00}^\Omega$ at $r=0$ as a function of the collision energy $\sqrt{s}$ along the freezeout line-2 in Table II as shown the main text, where two different angular velocities are chosen.}
\label{fig:omega-dep}
\end{figure}

In Fig.~\ref{fig:omega-dep} we investigate the dependence of  thermodynamic spin correlation $\delta\rho_{00}^\Omega$ on the angular velocity, where $r=0$ is chosen. It is found that the height of the peak increases with the angular velocity, similar as the $r$-dependence shown in Fig.~\ref{fig:oam}. In the same way, the increasing angular velocity not only enhances the quark-antiquark spin correlation directly, but also moves the CEP closer to the freezeout line. Furthermore, the angular-velocity dependence far away from the CEP at finite $r$ can also be found in Fig.~\ref{fig:r-dep}, where the influence from the change of CEP location is negligible. One finds that far away from the CEP the quark-antiquark spin correlation is dependent on the angular velocity approximately through $\sim\Omega^2/T^2$ as expected \cite{Yang:2017sdk}.

%
%

\section{Kurtosis of quark spin}

\begin{figure}[t]
\includegraphics[width=0.5\textwidth]{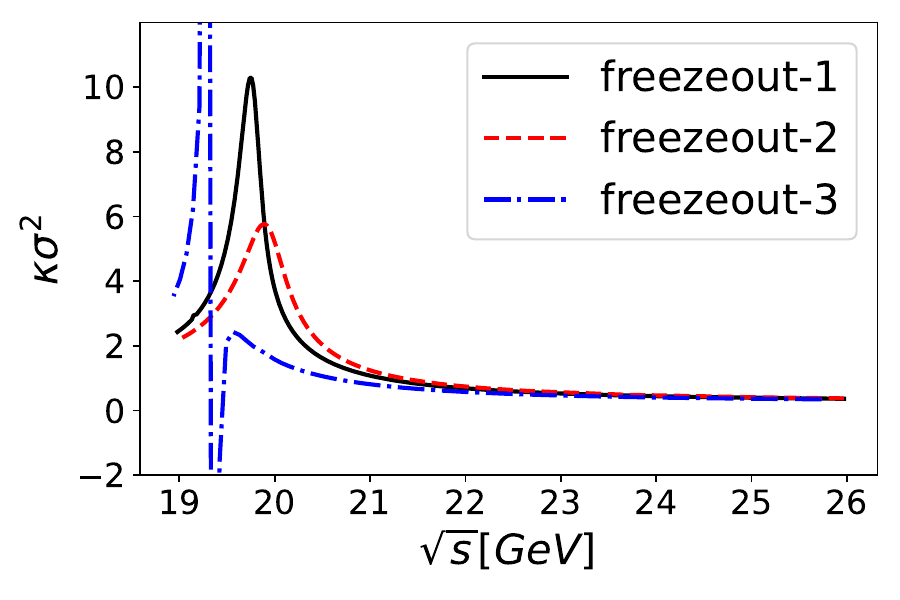}
\caption{Kurtosis of quark spin at $r=0$ on the three different freezeout lines.}
\label{fig:kurtosis_freezeout}
\end{figure}

We can also calculate the kurtosis of quark spin by the fourth derivative.
\begin{equation}
 	C_4=-\frac{V}{T}\frac{\partial^4 V^0_{\mathrm{eff}}}{\partial(\frac{\Omega}{T})^4}
 \end{equation}
  The numerical results of the kurtosis along the hypothetical freezeout lines defined in Table II is shown in Fig.~\ref{fig:kurtosis_freezeout}. Since higher cumulants are more sensitive to critical fluctuation, the peak structure is more obvious than that in quark-antiquark correlation. Our argument in the main text can also be applied to the kurtosis, which leads to the conclusion that  the negative kurtosis region of the whole system becomes wider. In experiment, this spin kurtosis can be measured by four $\Lambda$ correlations.
%
%


\section{PNJL model} 

We also check the case including the gluon field by performing the calculations once again in the PNJL model. The effective potential in the PNJL model at $r=0$ reads
\begin{equation}
\begin{split}
\label{eq:pnjl}
&V_{\mathrm{PNJL}}(\Omega_q^s,\Omega_{\bar q}^s, \mu_q,\mu_{\bar q};r=0)\\
=&\frac{[m-m_0]^2}{4G_{\mathrm{PNJL}}}-2N_f\int_0^\Lambda\frac{\ud^3p}{(2\pi)^3}3\varepsilon_p\\
&-N_f\int_0^\infty\frac{\ud^3p}{(2\pi)^3}\Big[T\ln(1+3\Phi\ue^{-(\varepsilon_p-\mu-\Omega_q^s/2-\mu_q)/T}+3\bar\Phi\ue^{-2(\varepsilon_p-\mu-\Omega_q^s/2-\mu_q)/T}+\ue^{-3(\varepsilon_p-\mu-\Omega_q^s/2-\mu_q)/T})\\
&+T\ln(1+3\Phi\ue^{-(\varepsilon_p-\mu+\Omega^s_q/2-\mu_q)/T}+3\bar\Phi\ue^{-2(\varepsilon_p-\mu+\Omega_q^s/2-\mu_q)/T}+\ue^{-3(\varepsilon_p-\mu+\Omega_q^s/2-\mu_q)/T})\\
&+T\ln(1+3\Phi\ue^{-(\varepsilon_p+\mu-\Omega_{\bar q}^s/2-\mu_{\bar q})/T}+3\bar\Phi\ue^{-2(\varepsilon_p+\mu-\Omega_{\bar q}^s/2-\mu_{\bar q})/T}+\ue^{-3(\varepsilon_p+\mu-\Omega_{\bar q}^s/2-\mu_{\bar q})/T})\\
&+T\ln(1+3\Phi\ue^{-(\varepsilon_p+\mu+\Omega_{\bar q}^s/2-\mu_{\bar q})/T}+3\bar\Phi\ue^{-2(\varepsilon_p+\mu+\Omega_{\bar q}^s/2-\mu_{\bar q})/T}+\ue^{-3(\varepsilon_p+\mu+\Omega_{\bar q}^s/2-\mu_{\bar q})/T})\Big]\\
&+T^4\bigg\{-\frac{1}{2}\Big[a_0+a_1(\frac{T_0}{T})+a_2(\frac{T_0}{T})^2\Big]\bar\Phi\Phi+b_3(\frac{T_0}{T})^3\ln\Big[1-6\bar\Phi\Phi+4(\bar\Phi^3+\Phi^3)-3(\bar\Phi\Phi)^2\Big]\bigg\},
\end{split}
\end{equation}
where $\Phi$ and $\bar\Phi$ are the Polyakov loop fields. The parameters we use are summarized in Table. \ref{table:para_pnjl}. The equations of motion are given by
\begin{equation}
\begin{split}
\frac{\partial V_{\mathrm{PNJL}}}{\partial m}=0, \quad \frac{\partial V_{\mathrm{PNJL}}}{\partial \Phi}=0, \quad \frac{\partial V_{\mathrm{PNJL}}}{\partial \bar\Phi}=0.
\end{split}
\end{equation}
%
\begin{figure}[t]
\includegraphics[width=0.5\textwidth]{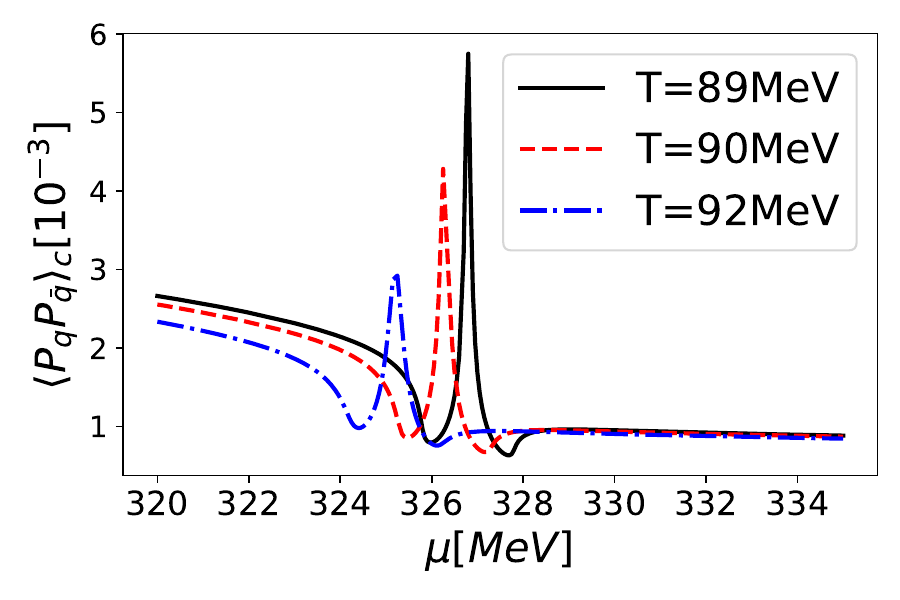}
\caption{Quark-antiquark spin correlation $\langle  P_{q}P_{\bar q}\rangle_{c}$  as a function of the chemical potential at $r=0$ and $\Omega=10$ MeV obtained in the PNJL model.}
\label{fig:pnjl}
\end{figure}
%
%
\begin{table}[t]
\caption{Parameter set used in the PNJL model \cite{Sun:2020bbn}.
}\label{table:para_pnjl}
\begin{ruledtabular}
\begin{tabular}{ccccccccc}
$\Lambda$ [MeV]&  $m_0$ [MeV] & {}$G_{\mathrm{PNJL}}\Lambda^2$ & $N_f$& $a_0$& $a_1$&$a_3$&$b_3$&$T_0$[MeV]\\
\colrule
$651$  & $5.5$ & $2.135$ & $2$& $3.51$&$-2.47$&$15.2$&$-1.75$&$210$\\ 
\end{tabular}
\end{ruledtabular}
\end{table}
%

It is straightforward to obtain the quark-antiquark spin correlation by taking derivative of Eq.~(\ref{eq:pnjl}). The spin correlations near the CEP are presented in Fig.~\ref{fig:pnjl}. In comparison to the lower panel of Fig. 1 in the main text, there is no qualitative difference between the NJL and PNJL results. In both cases, we observe a sharp peak as approaching the CEP. Since the peak structure of the quark-antiquark correlation is a consequence of critical fluctuations near the CEP, it should be robust regardless of the gluon background fields. 

It should be mentioned that, for the Polyakov loop potential we do not include an explicit rotation dependence. Such a dependence, i.e., the polarized Polyakov loop potential induced by the rotation, was studied and considered based on lattice results in \cite{Sun:2024anu}. Note that in the PNJL model the explicit rotation dependence of the Polyakov loop potential is only related to the angular momentum of gluon field, which is irrelevant when we take derivative w.r.t. the angular velocity coupled to quark spin, especially at $r \sim 0$. Thus, it is expected that this effect merely affects the location of CEP, and the peak structure of the spin correlation near the CEP is a robust conclusion.


\end{document}